# A Novel Decision Ensemble Framework: Customized Attention-BiLSTM and XGBoost for Speculative Stock Price Forecasting

Riaz Ud Din[1, 2], Salman Ahmed[2], Saddam Hussain Khan[1]*

[1]Artificial Intelligence Lab, Department of Computer Systems Engineering, University of Engineering and Applied Sciences (UEAS), Swat, Pakistan

[2]Department of Computer Systems Engineering, University of Engineering and Technology (UET), Peshawar, Pakistan

**Email:** saddamhkhan@ueas.edu.pk

## Abstract

Forecasting speculative stock prices is essential for effective investment risk management that drives the need for the development of innovative algorithms. However, the speculative nature, volatility, and complex sequential dependencies within financial markets present inherent challenges which necessitate advanced techniques. This paper proposes a novel framework, CAB-XDE (customized attention BiLSTM-XGB decision ensemble), for predicting the daily closing price of speculative stock Bitcoin-USD (BTC-USD). CAB-XDE framework integrates a customized bi-directional long short-term memory (BiLSTM) with the attention mechanism and the XGBoost algorithm. The customized BiLSTM leverages its learning capabilities to capture the complex sequential dependencies and speculative market trends. Additionally, the new attention mechanism dynamically assigns weights to influential features, thereby enhancing interpretability, and optimizing effective cost measures and volatility forecasting. Moreover, XGBoost handles nonlinear relationships and contributes to the proposed CAB-XDE framework's robustness. Additionally, the weight determination theory-error reciprocal method further refines predictions. This refinement is achieved by iteratively adjusting model weights. It is based on discrepancies between theoretical expectations and actual errors in individual customized attention BiLSTM and XGBoost models to enhance performance. Finally, the predictions from both XGBoost and customized attention BiLSTM models are concatenated to achieve diverse prediction space and are provided to the ensemble classifier to enhance the generalization capabilities of the proposed CAB-XDE framework. Empirical validation of the proposed CAB-XDE framework involves its application to the volatile Bitcoin market, using a dataset sourced from Yahoo Finance (BitCoin-USD, 10/01/2014 to 01/08/2023). The proposed CAB-XDE framework outperforms state-of-the-art models with a MAPE of 0.0037, MAE of 84.40, and RMSE of 106.14. The proposed CAB-XDE framework presents a technique for informed decision-making in dynamic financial landscapes, demonstrating effectiveness in handling the complexities of BTC-USD data.

# 1. Introduction

In today's information-driven era, the stock market remains a global economic epicenter with far-reaching impacts on commerce, industry, and society. The complexities of financial markets necessitate predictive techniques that play a critical role in mitigating risks and optimizing investment choices [1]. The diverse stock market faces challenges due to the complex contrast between conventional and speculative stocks, especially cryptocurrencies like Bitcoin. The speculative nature of Bitcoin which is driven by volatility and market sentiment introduces complexities demanding innovative predictive techniques.

However, challenges arise when applying these innovations to speculative stocks like Bitcoin. The distinctive characteristics of cryptocurrencies necessitate a reevaluation of existing predictive models. . As of February 2023, the cryptocurrency landscape boasts over 22,000 diverse tokens, collectively valued at over $1 trillion. With its $475 billion valuation and 42% market share, Bitcoin, which was founded in 2009, is the most powerful of them. Predicting Bitcoin prices poses a crucial yet challenging task for investors due to the absence of a benchmark futures market, the decentralized nature of Bitcoin transactions, and the difficulty in identifying factors influencing its price movements [2]. However, both traditional and speculative stocks share commonalities in price movements influenced by factors like supply and demand, investor sentiment, and global economic situations [3].

Traditional stock price prediction methods rely on fundamental and technical analysis and exhibit limitations hindering adaptability to the evolving dynamics of stock and cryptocurrency markets [4] [5]. A paradigm shift is witnessed with the integration of algorithmic trading, artificial intelligence (AI), and big data analytics, wherein machine learning (ML) and deep learning (DL) emerge as transformative contributors, offering substantial improvements over traditional methods [6]. DL techniques have demonstrated effectiveness in capturing these complicated relationships and predicting price movements in both traditional and speculative stocks [7]. Moreover, DL has already played a vital role in cancer diagnosis [8], [9], viral infection detection [10]–[12], cyber security [13], [14], intelligent transportation [15], [16], etc. DL, especially Recurrent Neural Networks (RNN) like LSTM offer significant feature engineering to capture long-range dependencies and temporal patterns in financial time series data [17]. Moreover, the effectiveness of combining LSTM with Convolutional Neural Networks (CNN) in addressing correlated data is evident, enhancing the accuracy of stock price predictions [18].

This research systematically integrates a Bitcoin dataset, moving beyond traditional stock market constraints. Bitcoin is characterized by heightened volatility and continuous global operation, making it a distinctive testing ground for the proposed CAB-XDE framework. This framework provides valuable insights that extend beyond Bitcoin. These insights have the potential to influence traditional stock prices.The following sections will detail the integration of BiLSTM models, attention mechanisms, and XGBoost presenting a novel approach to enhance price prediction performance for both traditional and speculative stocks. Their ability to model complex relationships and capture long-range dependencies positions them as potent tools for predictive analytics in financial markets.

**Our Contributions are as follows:**

1. A novel framework CAB-XDE is proposed for bitcoin daily price prediction. The proposed CAB-XDE seamlessly ensembles the learning capabilities of a customized BiLSTM, insights from a new attention mechanism, and the improved predictive proficiencies of the modified XGBoost algorithm.

2. The customized BiLSTM excels in capturing complex sequential dependencies and recognizing trends in speculative market trends. The incorporation of the new attention mechanism dynamically assigns weights and spotlights influential features. Additionally, the inclusion of XGBoost, renowned for its proficiency in handling nonlinear relationships, contributes to the robustness of the framework and ensuring more reliable predictions.

3. The weight determination theory-error reciprocal method is employed to further refine predictions from the individual new attention customized BiLSTM and XGBoost models. This iterative process is guided by disparities between theoretical expectations and actual errors to further enhance the model's performance.

4. Finally, the prediction is achieved through an ensemble learning approach, integrating individual predictions from both XGBoost and new attention customized BiLSTM models. This systematic integration enhances the diversity of predicted prices, thereby contributing to the improved generalization capabilities of the proposed CAB-XDE framework. Empirical validation of the proposed CAB-XDE framework on the BTC-USD dataset reveals better performance. The proposed CAB-XDE framework effectively addresses complexities and volatility in bitcoin prices, optimizing cost measures and outperforming recent state-of-the-art techniques.

The forthcoming research will be organized as follows: the literature review will be the main emphasis of Section 2, which comes next. In Section 3, the proposed CAB-XED framework is thoroughly explained. Section 4 will delve into the details of the experimental setup. The analysis of the proposed CAB-XDE framework and a comparison with state-of-the-art models will be presented in Section 5 through results and discussion. Finally, Section 6 will encapsulate the study's conclusion and potential future directions.

## 2. Literature View

Traditionally, stock market prediction relied on fundamental and technical analyses, but recent years have introduced a transformative era with the integration of sophisticated ML and deep learning techniques [19]. DL emerges as a potent tool, demonstrating its efficacy in diverse applications like financial analysis, emerging infectious disease, security, transportation, etc [20]–[25]. The nonlinear, noisy, and chaotic nature of stock market data has prompted the adoption of artificial intelligence (AI)-based techniques, particularly ML, with methods such as Support Vector Regression (SVR) and Artificial Neural Networks (ANN) gaining popularity. However, the sensitivity of these methods to data quality and quantity remain a challenge [26], [27]. Advancements in DL, a subfield of ML, has played a pivotal role in revolutionizing stock market forecasting. CNN and LSTM are the type of DL, have demonstrated effectiveness in extracting complex patterns from large datasets [28]. The evolution of LSTM into the BiLSTM model has shown remarkable ability in financial series prediction [29]. However, the inherent unpredictability of stock market data often necessitates a multi-pronged approach, as singular deep learning models can fall short. Moreover, the quality of data samples significantly influences model performance [30].

Notably, the hybrid decomposition-reconstruction model, incorporating RNN with gated recurrent units (GRUs), variational modal decomposition (VMD), and sample entropy (SE), has shown significant effectiveness [31]–[34]. The success of hybrid models is further exemplified by the EMD-BiLSTM model, which achieves a remarkable enhancement in forecasting accuracy, and the integration of an attention mechanism into the BiLSTM model, significantly elevating accuracy from 58.50% to 71.26% [35], [36]. These advancements underscore the potential of innovative model architectures and integration strategies in refining forecasting methodologies.

Transitioning to speculative stocks, the study focuses on Bitcoin price prediction, highlighting the performance of the LSTM-BTC model and its uncertainties regarding future data and generalizability to other cryptocurrencies [37]. Comparative studies using random forest regression and LSTM reveal insights into refining Bitcoin price prediction [38]. The importance of sample dimensions in ML algorithms for accurate Bitcoin price prediction is emphasized, with Logistic Regression and XGBoost achieving specific accuracies [39]. Exploration of optimized deep learning models for various cryptocurrencies and the evaluation of RNN (LSTM, GRU, BiLSTM) for major cryptocurrencies reveal promising avenues for accurate cryptocurrency price predictions [40]. The literature suggests a paradigm shift toward combination methods, emphasizing the necessity for diverse algorithms to address the evolving dynamics of financial markets [41].

Addressing the limitations of prior research, particularly those associated with daily trading volume and closing prices, this study proposes the adoption of ensemble techniques. The challenges inherent in daily trading volume and closing prices, encompassing volatility, non-linearity, external dependencies, noise, lack of clear patterns, and data quality issues, accentuate the necessity for resilient ensemble approaches to adeptly navigate the intricacies of financial market data. Employing ensemble methods, such as tagging or boosting, to combine multiple models or predictions emerges as a strategic measure to reduce the risk of overfitting and improve overall performance by harnessing the strengths inherent in diverse models.

While the BiLSTM has demonstrated its effectiveness in sequence modeling, it encounters difficulties when employed for stock price prediction. A significant issue lies in its uniform weight assignment to input features, neglecting the diverse levels of importance that impact stock prices. In response to this concern, the incorporation of the attention mechanism dynamically assigns weights, highlighting the most influential features [30]. Although the attention mechanism greatly increases the BiLSTM performance, it is important to recognize that overfitting and the black-box aspect of neural networks remain fundamental BiLSTM constraints. To navigate these challenges, a combination with the XGBoost model is introduced. XGBoost excels in managing intricate nonlinear relationships, ensuring robustness and delivering enhanced interpretability compared to neural networks [26]. This strategic combination results in the development of the the proposed CAB-XDE framework. This novel approach undergoes empirical validation using Bitcoin data, representing a significant step in addressing challenges unique to speculative stocks within dynamic financial markets.

## 3. Proposed Framework CAB-XDE

This paper introduces a novel methodology for enhancing the predictive performance of the proposed CAB-XDE framework. This methodology leverages the combined strengths of three distinct techniques: a customized BiLSTM with attention mechanism, a robust XGBoost model, and a strategic error reciprocal weighting scheme. This section details the individual model training processes, the error reciprocal weighting calculation, and the ensemble prediction generation procedure. Moreover, the proposed CAD-XDE framework has beem employed on bitcoin strengtht dataset and compared with the best state of the art techniques, graphical overview is shown in Figure 1. Additionally, scaling has been utilized as preprocessing to normalize the whole dataset.

### 3.1 Ensemble New attention customized BiLSTM and XGBoost

The proposed CAB-XDE framework prediction model surpasses the limitations of traditional statistical methods (e.g., linear regression, ARIMA, SMA) by leveraging the integration between DL and ensemble techniques. For capturing complex sequential dependencies and identifying market trends, the CAB-XDE employs a new attention-Improved customized BiLSTM. This innovative architecture effectively analyzes temporal relationships within the data. Additionally, an integrated XGBoost module enhances CAD-XDE stability and generalizability by tackling nonlinearities and preventing overfitting. Moreover, the error reciprocal method analyzes the predictions of individual new attention customized BiLSTM and XGBoost and assigns dynamic weights, favoring models with lower error rates. This weighted integration mitigates intrinsic limitations of individual models and offers a robust framework for tackling the multifaceted challenges of the dynamic and speculative Bitcoin market. The proposed CAB-XDE framework general and detailed overview is visually represented in Figure 2 and Figure 3 capitalizing on the strengths of both individual components, provides a powerful and comprehensive foundation for better Bitcoin price forecasting.

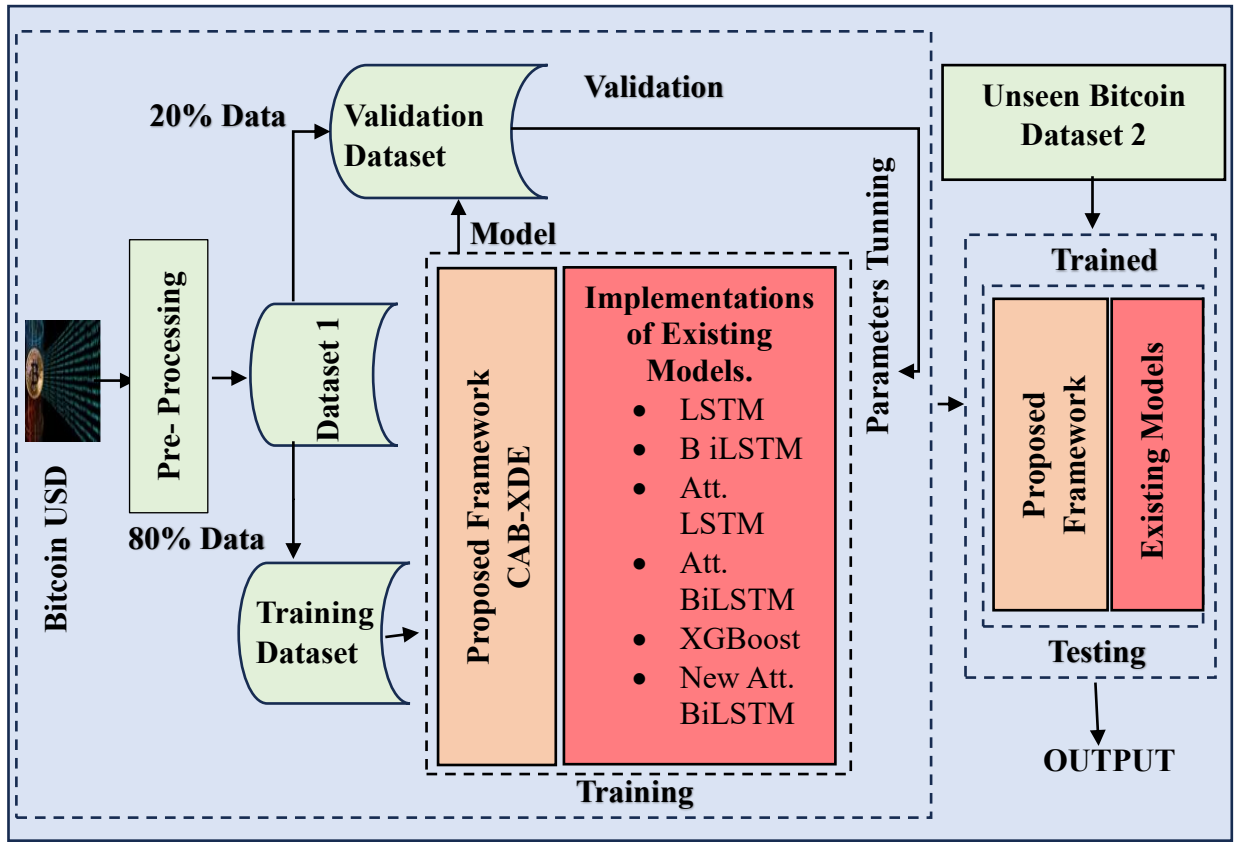

*Figure 1. Graphical overview of the proposed study.*

## 3.2 Weighting Method

The error reciprocal method plays a pivotal role in elevating prediction performance. By assigning inverse weights to errors, this method accentuates instances where the model makes larger errors, focusing on significant deviations between predicted values and actual outcomes. It is useful for determining which adjustments to prioritize for cases with large deviations, which improves the model's capacity to identify and address significant differences in its predictions [42].

This paper employs the error reciprocal method for weight assignment to enhance the predictive accuracy of proposed CAB-XDE framework. The method calculates weights based on error outcomes from the primary evaluation metric, MAPE. Assigning larger weights to models with smaller errors within this composite framework markedly reduces the overall prediction error. This weight determination is expressed through the following formula:

$$P_{bl(t)} = W_{bl} \, P_{bl(t)} \ + W_{xg} \, P_{xg(t)}, \ t= 1, 2, \ldots, n \tag{1}$$

$$W_{bl(t)} \ = \frac{Exg}{Ebl + Exg} \tag{2}$$

$$W_{xg(t)} \ = \frac{Ebl}{Ebl + Exg} \tag{3}$$

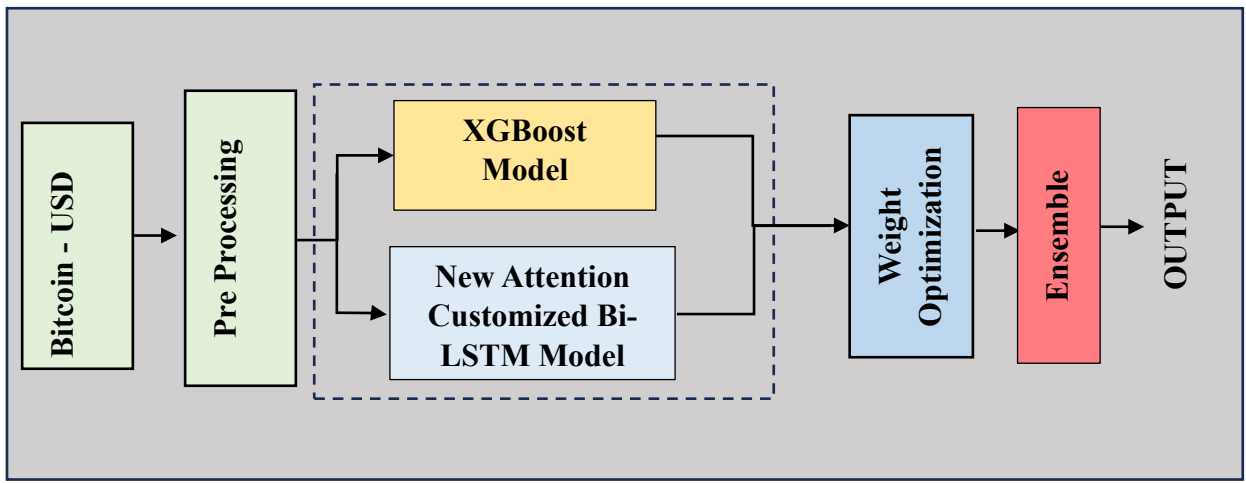

*Figure 2. Proposed bitcoin prediction CAB-XDE framework.*

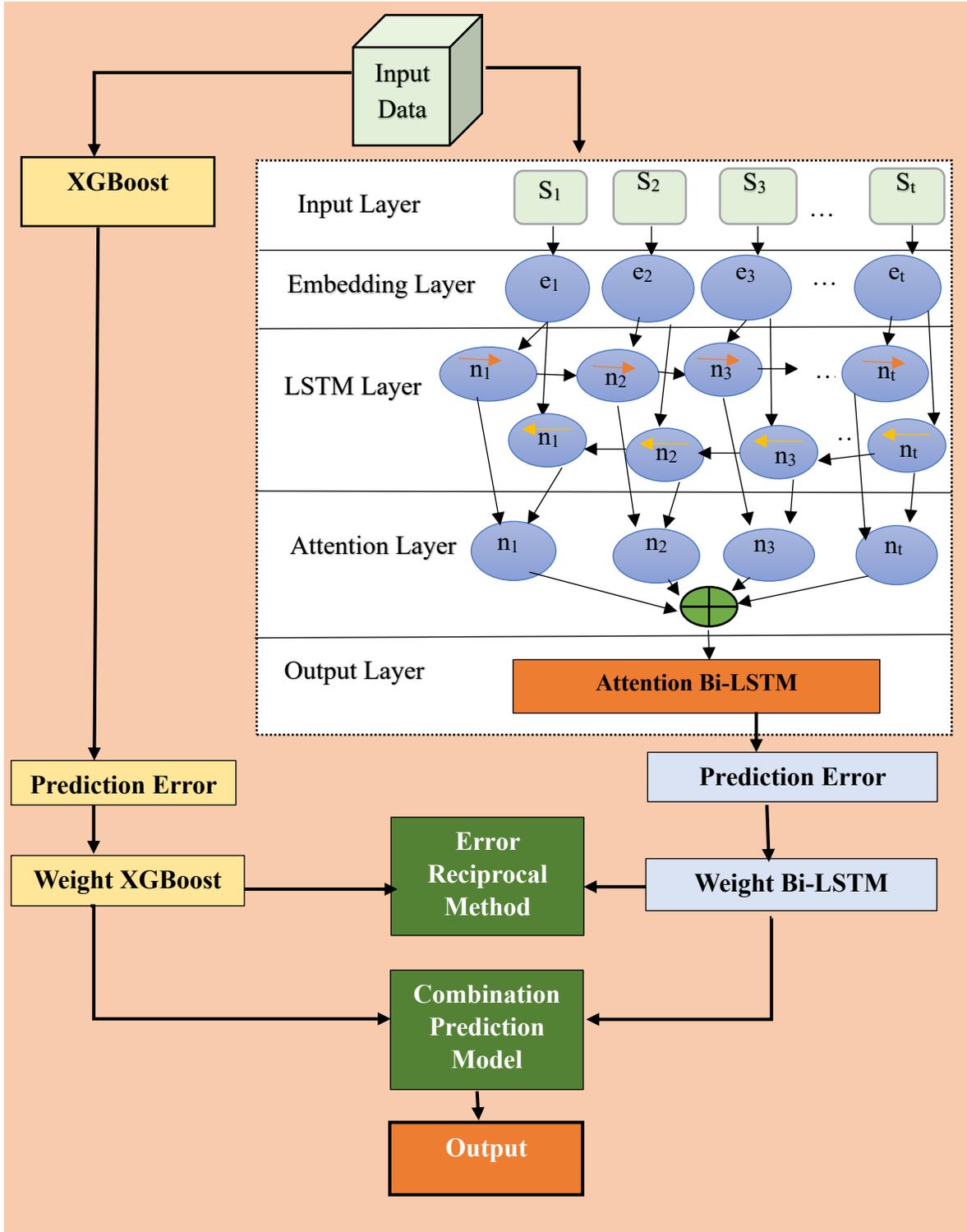

*Figure 3. Proposed CAB-XDE detailed framework.*

$\mathbf{W_{bl(t)}}$ and $\mathbf{P_{bl(t)}}$ denote the weight values and predicted values of new attention customized BiLSTM respectively. While $\mathbf{W_{xg(t)}}$ and $\mathbf{P_{xg(t)}}$ represent predicted values of XGBoost respectively.

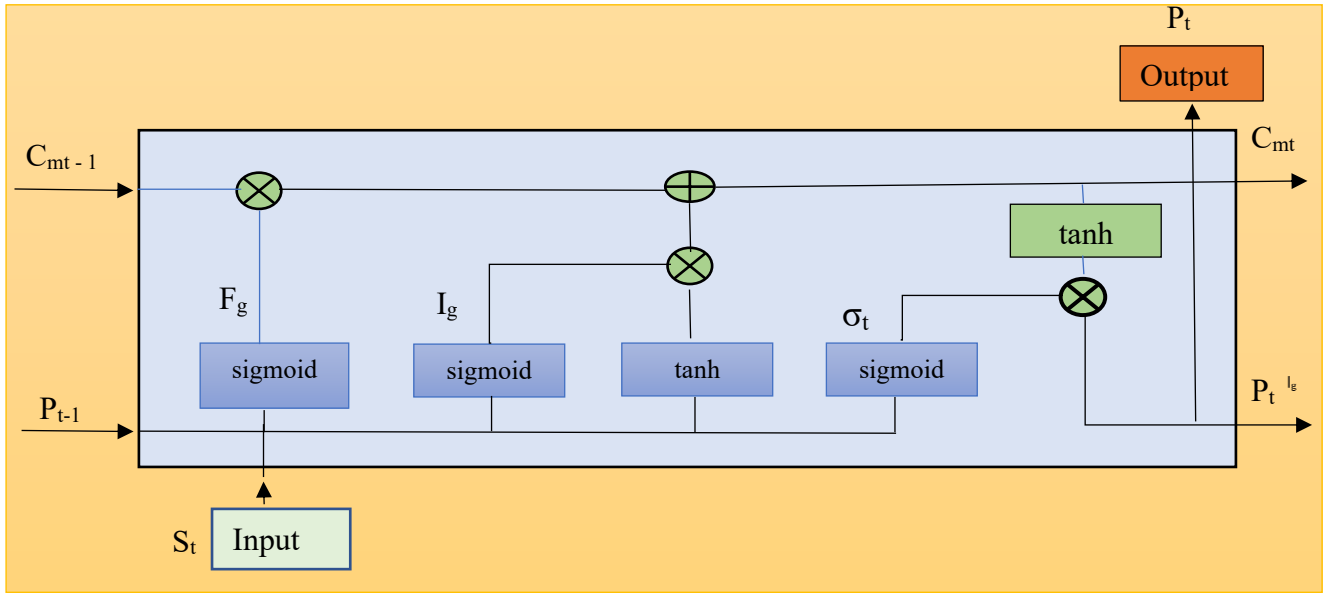

*Figure 4. Gating mechanism of LSTM.*

The weight calculations are based on formulas (2) and (3), where the error values for the new attention BiLSTM and XGBoost are represented by the variables **Ebl** and **Exg** respectively.

### 3.3 BiLSTM Prediction Model

This paper focuses on predicting Bitcoin (BTC) prices proposing a novel CAB-XDE framework. The chosen dataset, sourced from Yahoo Finance and encompassing BTC-USD exchange rates, presented inherent limitations necessitating a unique approach for better forecasting. These challenges include the intricate  dynamic pattens, market volatility, and the sequential nature inherent in bitcoin price [43]. In response to these complexities, the proposed CAB-XDE framework employs a new attention Cutomized BiLSTM neural networks. BiLSTM is chosen for its unique ability to effectively capture complex sequential dependencies and determine patterns in sequential data, aligning seamlessly with the dynamic nature of Bitcoin price [40].

Recurrent neural networks (RNNs) serve as powerful tools for modeling sequential data; however, their effectiveness hinges significantly on the selection of time steps. This is because the gradient backpropagation algorithm relies on the chain rule to compute the derivatives of the loss function concerning the network parameters. When the time steps are very large, the gradients can become vanishingly small, hindering the learning process. Conversely, when the time steps are very small, the gradients can become very large that can lead to exploding gradient, causing instability and hindering convergence [44]. In order to overcome this, the LSTM model—shown in Figure 4, below,—introduces a gating mechanism made up of input, output, and forgetting gates. The information that is exposed to successive layers is controlled by the output gate, the forget gate controls the information that is removed from the concealed state, and the input gate regulates the infusion of new information. The limitations of typical RNNs are overcome by LSTM networks, which can successfully learn long-term dependencies inside sequences due to the complex interplay between the gates. The hidden state, which acts as a link between the previous time step

$H_{t-1}$ and the current time step $X_t$, provides the input for the LSTM gate. Subsequently, a fully connected layer is used to calculate the LSTM's output [30].

Input Gate: $I_g = \sigma(S_t WM_{xn} + Ph_{t-1} W_{nIg} + d_n)$      (4)

Forget Gate: $F_g = \sigma(S_t W_{xf} + Ph_{t-1} W_{nFg} + d_f)$      (5)

Gated unit: $C_m = \tanh(S_t WM_{xc} + Ph_{t-1} W_{nCm} + d_c)$      (6)

$C_{mt} = F_g \odot C_{m(t-1)} + I_g \odot C_{mt}$      (7)

Output gate: $O_g = \sigma(S_t W_{og} + Ph_{t-1} W_{nOg} + d_o)$      (8)

**'$Ph_{t-1}$'** signifies the hidden state from the preceding time step, the small set input at a given time step **'t'** is indicated by **'$S_t$'**, and the number of hidden state is indicated by **'hs'**. **'$WM_{xn}$'** & **'$W_{nIg}$'** stand for the weight matrices of the input gate, while **'$\sigma$'** denotes the sigmoidal function. Additionally, the term **'dn'** serves as an offset term of the input gate. In addition, the weight matrices assigned to the forgetting gate are denoted by the symbols **'$W_{xf}$'** and **'$W_{nFg}$'** in this architecture, while the associated offset term is represented by the symbol **'df'**. **'$WM_{xc}$'** and **'$W_{nCm}$'** are used as weight matrices for the gated unit, while **'dc'** stands for the related offset term. The term **'$C_m$'** is introduced to describe the candidate memory cells. Furthermore, **'$C_{mt}$'** designates the cell state for the current time step, while **'$C_{m(t-1)}$'** represents the cell state for the previous time step. Finally, **'do'** is the offset term that corresponds to the defined weight matrices **'$W_{og}$'**, which are linked to the output gate. The activation function used for regulating information flow in the hidden state involves the multiplication of elements (x) and employs the tanh function, known for its value range of [-1,1].

In contrast to the conventional LSTM, the BiLSTM approach seamlessly integrates forward and backward LSTMs. The proposed CAB-XDE framework employs a specialized feature extraction method, leveraging the comprehensive information within the data to capture insights from both forward and backward perspectives. The outcomes of this two-way extraction are harmoniously combined and summarized in two dimensions. By strategically merging the data, the inherent influence of the order of inputs on a single LSTM is mitigated, enhancing the overall comprehensiveness of the outputs. The resulting enriched output, derived from both perspectives, is denoted as

$Fo = Og(x) \tan h(Cm)$      (9)

ensuring a comprehensive representation of features, Figure 5 illustrates the components' connection topology of LSTM [30]. This methodological foundation serves as a robust foundation, effectively tackling challenges related to gradient instability and sequential feature extraction in speculative stock price prediction endeavors.

## 3.4 attention mechanism

The BTC-USD dataset often involves subtle features and patterns that may carry varying degrees of importance. The introduction of the new attention mechanism to BiLSTM dynamically assigns weights to influential features, thereby enhancing interpretability and optimizing cost measures and volatility predictions [45]. The foundational concept behind the attention mechanism is inspired by human attention dynamics. In human information processing, attention is selectively focused on key elements rather than uniformly distributed across all information. Integrating the attention mechanism into prediction models mirrors this cognitive approach, enabling the assignment of distinct weights to data. This dynamic allocation mitigates the undue influence of certain input data on the output, amplifying the significance of pivotal information. The attention structure, illustrated in Figure 6, describes how this mechanism refines the model's focus. Within the customized BiLSTM model, two pivotal attention strategies emerge. Notably, the attention gate replaces the conventional forget gate seen in traditional LSTM models. This gate exclusively attends to historical cell states, untied from the current input, leading to a notable reduction in overall training parameters. Additionally, the BiLSTM model employs an attention weighting method on the model's output. This strategic application allows for the precise identification and utilization of the most crucial and influential information [36].

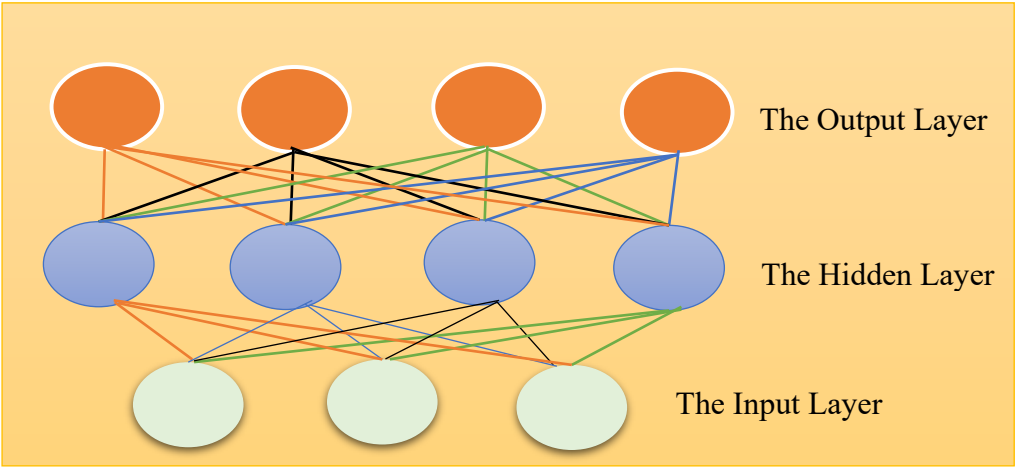

*Figure 5. Component connection topology of LSTM.*

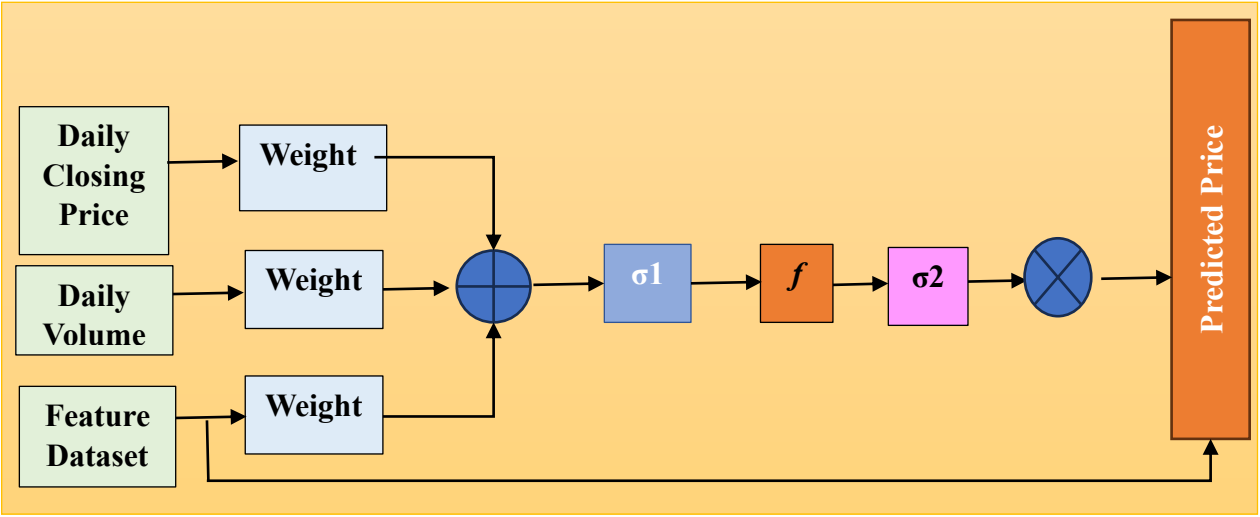

*Figure 6. Structure of attention mechanism.*

## 3.5 XGBoost Forecasting Model

Despite the promising performance of the new attention customized BiLSTM in Bitcoin price prediction, it is not without its inherent limitations. In response to these challenges, XGBoost is used for robust gradient-boosted decision tree implementation. XGBoost excels in handling diverse data types, managing nonlinear relationships, and preventing overfitting. XGBoost ensures model stability and generalizability thereby addressing the limitations of over-reliance on sequential information and enhancing model interpretability. Furthermore, XGBoost's unique ability to combine weak learners facilitates the capture of complex dynamic patterns, contributing to more accurate predictions even during significant market shifts [46]. XGBoost represents a refined instantiation of the gradient boosting decision tree paradigm, distinguished by its ability to improve predictive speed and efficiency. XGBoost employs a meticulously crafted decision tree constructed iteratively by adding trees and progressively partitioning features. This sophisticated approach contributes to the model's enhanced performance in capturing complex patterns within the Bitcoin price data [2]. The residual error from the previous prediction is represented by creating a new function called $f(\mathrm{x})$ as part of this incremental integration process. After all of the k trees in the training process have been used, each tree converges to a different leaf node, and every leaf node has a different score. The ultimate forecast for a specific sample is then obtained by adding up the ratings linked to every contributing tree. The XGBoost model's formal statement is as follows:

$$\hat{p}_i = \sum_{j=1}^{n} w_j p_{ij} \tag{10}$$

Herein, forecast value is indicated by $\hat{p}_i$, sample data is indicated by $p_{ij}$. While n indicates the total number of trees and $w_j$ indicates the weight.

A new tree is slowly added to the existing tree structure in each iteration cycle to model the residual disparity between the results and the predictions of the previous tree. The following is an explanation of the iterative process:

$$\hat{p}_i^{(0)} = 0 \tag{11}$$

$$\hat{p}_i^{(1)} = f_1(p_i) = \hat{p}_i^{(0)} + f_1(p_i)$$

$$\hat{p}_i^{(2)} = f_1(p_i) + f_2(p_i) = \hat{p}_i^{(1)} + f_2(p_i)$$

$$\ldots$$

$$\hat{p}_i^{(t)} = \sum_{k=1}^{t} f_k(p_i) = \hat{p}_i^{(t-1)} + f_t(p_i)$$

$\hat{p}_i^{(t)}$ represents the model after $t$ training rounds, $\hat{p}_i^{(t-1)}$ signifies the retained function from the earlier round, whereas $f_t(p_i)$ is the recently introduced function. Objective function of XGBoost is stated as follows:

$$Obj(t) = \sum_{i=1}^{n} l\left(p_i, \hat{p}_i^{(t)}\right) + \sum_{i=1}^{t} \Omega(f_i)$$

$$= \sum_{i=1}^{n} l\left(p_i, \hat{p}_i^{(t-1)} + f_t(p_i)\right) + \Omega(f_t) \tag{12}$$

$$\Omega f(t) = \gamma L + \frac{1}{2}\lambda \sum_{j=1}^{T} l_j^2 \tag{13}$$

It is important to find $f_t$ that minimizes the chosen target function. Equation (12) has a regularization term called $\sum_{i=1}^{t} \Omega(f_i)$, which affects the objective function's tree's complexity. Notably, improved generalization ability and decreased complexity are correlated with a lower value of $\Omega(f_t)$. Equation (13) shows that L is the number of leaf nodes, $l$ is the score awarded to a leaf node, $\gamma$ controls the number of leaf nodes, and $\lambda$ limits the scores of leaf nodes to avoid unnecessarily high values.

Second-order expansion of Taylor at $f_t$=0 is used to determine $f_t$ that minimizes the goal function. Aapproximation resulting from the objective function is expressed as:

$$\tau^{(t)} \simeq \sum_{i=1}^{n} \left[ l\left(p_i, \hat{p}_i^{(t-1)}\right) + g_i f_t(p_i) + \frac{1}{2} h_i f_t^2(p) \right] + \Omega(f_t) \tag{14}$$

Here, $g_i = \partial_{\hat{p}}(t-1)l\left(p_i, \hat{p}^{(t-1)}\right)$, and $h_i = \partial_{\hat{p}^{(t-1)}}^2 l\left(p_i, \hat{p}^{(t-1)}\right)$ is the first and second derivatives respectively.

These elements are immediately eliminated because the residual error of p and the prediction scores from the original t-1 tree have no bearing on the optimization of the objective function. As a result, more simplification of the goal function produces:

$$\tilde{\tau}^{(t)} = \sum_{i=1}^{n} \left[ g_i f_t(p) + \frac{1}{2} h_i f_t^2(p_i) \right] + \Omega(f_t) \tag{15}$$

Equation (15) aggregates the values of the loss function for every sample, hence simplifying the objective function. Then, in order to simplify and rephrase the goal function, samples belonging to the same leaf node are rearranged using Equation (16). The steps involved are as follows:

$$Obj^{(t)} \simeq \sum_{i=1}^{n} \left[ g_i f_t(p_i) + \frac{1}{2} h_i f_t^2(p_i) \right] + \Omega(f_t)$$

$$= \sum_{i=1}^{n} \left[ g_i l_{q_{(p_i)}} + \frac{1}{2} h_i l_{q_{(p_i)}}^2 \right] + \gamma L + \lambda \frac{1}{2} \sum_{j=1}^{L} l_j^2 \tag{16}$$

$$= \sum_{j=1}^{L} \left[ \left(\sum_{i \in I_j} g_i\right) l_j + \frac{1}{2}\left(\sum_{i \in I_j} h_i + \lambda\right) l_j^2 \right] + \gamma L$$

Hence, by reformulating the previously provided formulas, the objective function transforms into a univariate quadratic function centered on the leaf node fraction ω. This transformation enables the utilization of the vertex formula to readily ascertain the optimal ω and its corresponding value for the objective function. The optimal ω j* and related objective function values for the recalibrated univariate quadratic function centered around the leaf node fraction ω can be obtained in the following manner:

$$l_j^* = -\frac{\sum_{i \in I_j} g_i}{\sum_{i \in I_j} h_i + \lambda} \tag{17}$$

$$Obj = -\frac{1}{2} \sum_{j=1}^{L} \frac{\left(\sum_{i \in I_j} g_i\right)^2}{\sum_{i \in I_j} h_i + \lambda} + \gamma L \tag{18}$$

$$v^n = \frac{v - v_{\min}}{v_{\max} - v_{\min}} \tag{19}$$

For effective computations and adherence to data input specifications, data normalization is a prerequisite. Cryptocurrency data is normalized using the formula [19], effectively confining the data within the [0, 1] range. In the above formula [19], $v$ and $v^n$ respectively denote the stock data value before and after normalization. $v_{\min}$ and $v_{\max}$ represent the minimum and maximum values of the stock data before normalization.

### 3.6 Individual Model Training and Testing

Initially, individual BiLSTM and XGBoost models undergo training and testing, and then their respective predictions are combined into an ensemble model as illustrated in Figure 7. The prediction method comprises four distinct stages, explained as follows:

**Stage 1 Data Preprocessing**: The initial stage entails comprehensive data preprocessing to enhance generalization. Selecting key features, like Daily Opening Price, Daily Maximum and Minimum Price, Daily Trading Volume, and Daily Closing Price, is crucial in this phase. This meticulous selection ensures that the data is well-prepared for subsequent analysis. The culmination of this stage involves the normalization of the data, ensuring it is appropriately scaled for further analysis.

**Stage 2 Single Model Prediction:** Enhancing the customized BiLSTM model involves integrating an attention mechanism that highlights volume and daily closing prices. This strategic emphasis signals an unsustainable trend, whereas a declining price coupled with increasing or stagnant volume suggests a transient downturn. This direct correlation between price and volume serves as a crucial market sustainability metric. The refinement of the customized BiLSTM model, achieved through the new attention mechanism, targets the removal of illogical variables. This targeted approach accentuates vital input data, ensuring a more comprehensive and refined model output. The groundwork for eventual convergence in Stage 3 is laid through the integration of independent forecasts from both the Single new attention customized BiLSTM and XGBoost models.

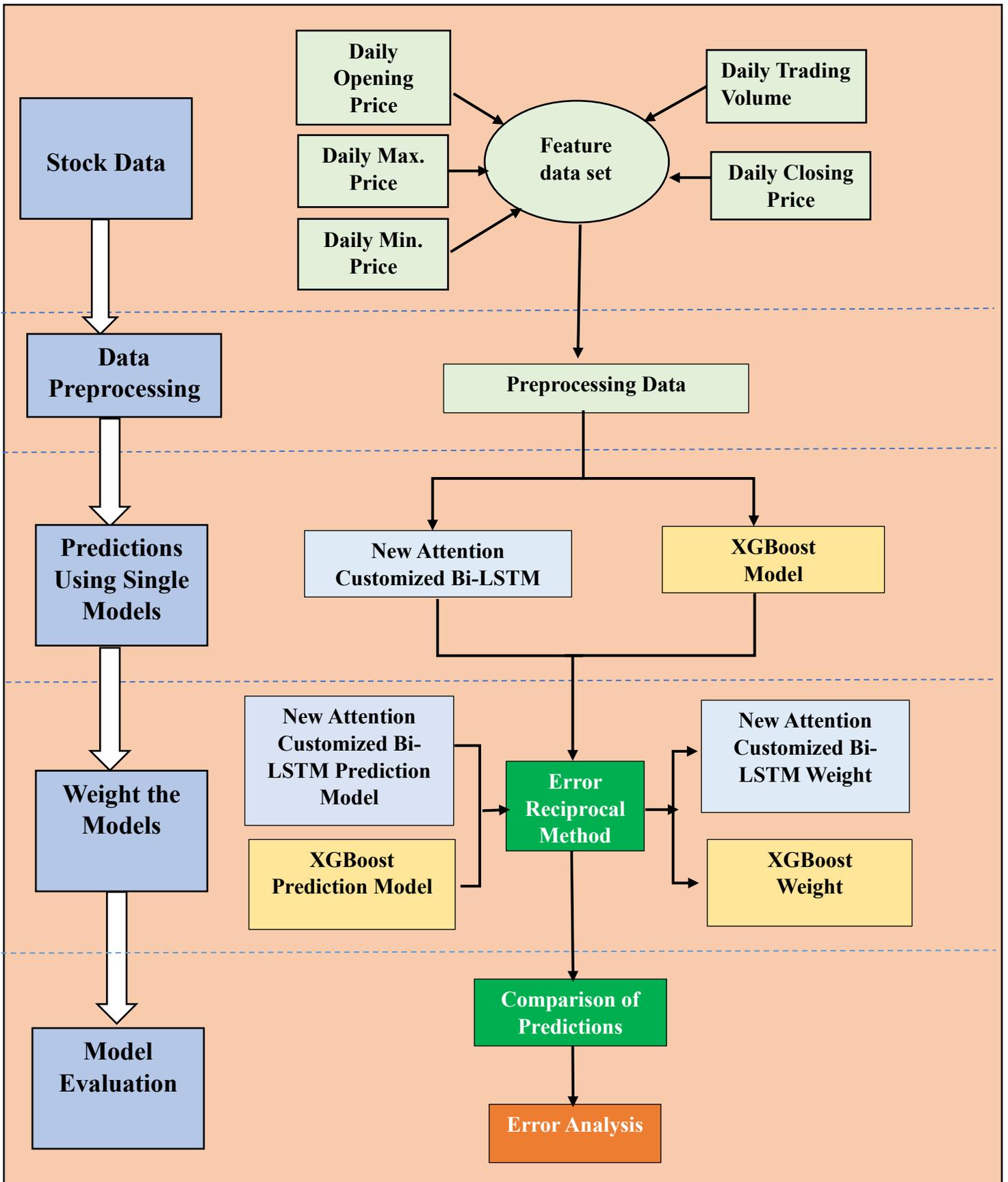

*Figure 7. Stock price prediction methodology in the proposed CAB-XED framework.*

**Stage 3 Assigning Weights to Models:** The new attention customized BiLSTM and XGBoost models are used to predict bitcoin price, and then the error reciprocal approach is employed to the predictions of the new attention customized BiLSTM and XGBoost models to assign weights depending on projected errors. This ensures that each model is appropriately weighted, leading to the development of the proposed CAB-XDE prediction framework.

**Stage 4 Assessing Prediction Performance:** Evaluating the effectiveness of the proposed CAB-XDE framework involves a comparison of prediction errors with six state-of-the-art models. The objective of this assessment is to measure the enhancement in Bitcoin price prediction performance attained through the proposed framework.

## 4. Experimental Setup

Assessing and scrutinizing forecasting technique's performance is accomplished through Python simulations. For thorough comparison, state-of-the-art models, including LSTM, attention-LSTM, BiLSTM, attention-BiLSTM, and new attention customized BiLSTM and XGBoost, are employed.

### 4.1 Datasets

Following a comprehensive assessment of data from diverse sources, Yahoo Finance stands out as the most current and comprehensive dataset for the CAB-XDE framework. Utilizing Date, Daily Opening Price, Daily Maximum Price, Daily Minimum Price, Daily Trading Volume, and Daily Closing Price, the developed approach undergoes evaluation based on these key data points. Structure of the Data is illustrated in Fig 9.

|  | Open | High | Low | Volume | Close |
|---|---|---|---|---|---|
| **Date** |  |  |  |  |  |
| **10/01/2014** | 387.427002 | 391.378998 | 380.779999 | 26229400.0 | 383.614990 |
| **10/02/2014** | 383.988007 | 385.497009 | 372.946014 | 21777700.0 | 375.071991 |
| **10/03/2014** | 375.181000 | 377.695007 | 357.859009 | 30901200.0 | 359.511993 |
| **10/04/2014** | 359.891998 | 364.487000 | 325.885986 | 47236500.0 | 328.865997 |
| **10/05/2014** | 328.915985 | 341.800995 | 289.295990 | 83308096.0 | 320.510010 |

*Figure 8. Structure of the data*

Following two datasets are employed for experimentation:

1. The first dataset (Dataset 1) covers the period from 10/01/2014 to 11/07/2022. Feature scaling is performed using MinMaxScaler (), with the target variable set as the Daily Close Price.

2. The second dataset (Dataset 2) spanning from 08/11/2022 to 01/08/2023, is acquired. Feature scaling is conducted using MinMaxScaler (), and the target variable is set as the Daily Close Price.

The new attention customized BiLSTM and XGBoost models are trained using 80% of the daily sampled historical bitcoin price data, while the remaining 20% is used for testing. The second dataset is exclusively reserved for evaluating the proposed CAB-XDE framework. Following this, a comparative analysis of the obtained results against the actual values is conducted, and a meticulous examination of the errors is undertaken.

## 4.2 Assessment Criteria

The primary goal is to assess the prediction performance of the developed CAB-XDE. Employing three commonly used statistical variables, the core evaluation metric for each prediction model is MAPE, complemented by auxiliary metrics like RMSE and MAE. The performance of the proposed CAB-XED framework is quantified through the following calculation procedures applied to these statistical measures:

$$X_{\text{MAPE}} = \frac{1}{n}\sum_{t=1}^{n}\frac{|r_t - \hat{p}_t|}{r_t} \qquad (20)$$

$$X_{\text{MAE}} = \frac{1}{n}\sum_{t=1}^{n}|r_t - \hat{p}_t| \qquad (21)$$

$$X_{\text{RMSE}} = \sqrt{\frac{1}{n}\sum_{t=1}^{n}(r_t - \hat{p}_t)^2} \qquad (22)$$

Where:

- n is the amount of bitcoin data,

- $r_t$ indicates the real stock price data, and

- $\hat{p}_t$ signifies the predicted stock price data.

## 4.3 Hardware setup

An AMD Ryzen 7 4800H with Radeon Graphics (16 CPUs), 3GHz, and 32GB RAM makes up the experimental hardware platform. Our developed framwork is implemented in Python, where XGBoost uses the py-xgboost framework and the new attention customized BiLSTM model uses the Keras DL framework.

## 4.4 Parameteric Configurations

### (i)New attention customized BiLSTM

The number of units, input feature dimension, number of unit layers, etc are important factors that affect the accuracy of the new attention customized BiLSTM model. For the new attention customized BiLSTM model's particular parameter configurations, see Table 1.

*Table 1. New attention customized BiLSTM parameter settings.*

| Model | Parameter | Description | Values | Explanation |
|---|---|---|---|---|
| New attention customized BiLSTM | Unit | The quantity of units | 99 | This parameter is crucial in determining the model's accuracy and needs to be optimized to reach its ideal value. |
| | time_step | Time step | 99 | Assessing the relationship between each input datum and the preceding sequential input data to determine their interdependence. |
| | num_layers | Number of layers in the unit. | 2 | Defaulting to 1 layer, setting it to 2 layers involves the second layer processing computational results from the first layer. |
| | batch_size | Width of the hidden layer. | 64 | The amount of data entered concurrently is determined by this parameter. The model can determine whether the input data is from the same batch thanks to its settings. |
| | Epochs | Amount of iterations | 64 | The computer's processing power determines the ideal number of iterations. |
| | Monitor Patience | Early Stopping | val_loss 10 | To prevent overfitting when the model's performance on a validation dataset ceases to improve. |
| | DropOut | Regularization technique. | 0.2 | A regularization method to avoid overfitting in neural networks. |

## (ii) **XGBoost Parameter Configurations**

The accuracy of XGBoost is predominantly influenced by several key factors. These include the iterative decision tree process, the number of decision trees, the choice of a weak evaluator, the XGBoost objective function, the progress of model training, the control of model complexity, parameters of regular terms, and the sample size of random sampling with replacement. Table 3 lists all of the XGBoost parameter configurations.

*Table 2. Parameter settings of XGBoost.*

| Model | Parameter | Description | Value | Explanation |
|-------|-----------|-------------|-------|-------------|
| XGBoost | reg | Objective | squared error | Instructing the algorithm to perform regression and minimize the mean squared error during the training process. |
| | n_estimators | The number of decision trees. | 100 | This parameter holds significant influence, capable of adjusting the model to its maximum extent in a single instance. |
| | max_depth | The maximum depth of decision tree. | 8 | The typical range is between 10 and 100, and adjustments can be made as needed, especially for larger sample sizes and feature quantities. |
| | silent | Model training progression. | 1 | In scenarios with extensive data and sluggish algorithmic speed, this parameter serves as a tool for monitoring the training progress. |
| | subsample | A random sampling place with a return sample size.. | 1 | This parameter controls the sample size during sampling.1 is the default value, signifying the extraction of 100% of the data at once, whereas a value of 0.1 implies the 10% extraction of the data at a time. |
| | eta | Iterative decision tree in the learning process. | 0.1 | The parameter η, referred to as the learning rate, denotes the step size of the iterative decision tree. Its significance lies in ensuring that each new tree makes an optimal contribution to the overall prediction effectiveness. |
| | booster | Selection of weak evaluator | gbtree | During the tree-building process, some trees are discarded, offering a superior overfitting function compared to gradient boosting trees. |
| | alpha | Parameters of regular terms | 10 | With larger values for alpha and lambda, imposing a heavier penalty, the proportion of regularization terms increases, resulting in a lower complexity for the model. |
| | gamma | Control of model complexity | 2 | Critical parameters for mitigating overfitting. |

## 5. Results

customized BiLSTM is employed to address the constraints of BTC-USD data, like complex dependencies, high volatility, and irregular updates. customized BiLSTM is specifically chosen for their adeptness at handling long-term dependencies and sequential learning inherent in time series data. However, the utilization of customized BiLSTM model introduces its set of challenges, notably susceptibility to overfitting, especially in the presence of noisy and limited financial data. The computational intensity and resource demands associated with training deep neural networks, including BiLSTMs, add an additional layer of complexity. Moreover, the interpretability of BiLSTMs is compromised due to their inherently black-box nature. In response to these limitations, new attention mechanism is introduced to improve interpretability by allowing selective focus on critical elements like daily closing pirce and daily volume within the input sequence. XGBoost is seamlessly integrated into the proposed CAB-XED framework to address the overfitting. XGBoost enchances the proposed CAB-XED framework with effective regularization techniques and adaptability to non-stationary data. XGBoost brings a complementary strength to the ensemble. The combination of new attention customized BiLSTM and XGBoost is then consolidated through an ensemble approach. This strategic integration capitalizes on the distinct architecture of both models, fostering a more resilient and accurate prediction model.

## 5.1 Discussion

Experiments are conducted to demonstrate the performance the proposed CAB-XED framework with Dataset 1 and Dataset 2, comprising daily Bitcoin price data. Initially, we individually trained and tested the new attention customized BiLSTM and XGBoost models using Dataset 1, revealing the prediction errors for each model. The results are depicted in Figure 9. Bitcoin Price Prediction Using BiLSTM with new attention mechanism Figure 9 and Figure 10, offering a visual representation of the prediction performance of each model. Upon computing errors, weights were assigned to the two aforementioned prediction models using the error reciprocal method and combined similarly. A higher weight was assigned to the model demonstrating a smaller error. Specifically, employing the primary evaluation index MAPE, the error reciprocal formula yielded weights of 0.4252 for the novel attention customized BiLSTM and 0.5748 for XGBoost. Following this determination, the ensemble of the new attention customized BiLSTM and XGBoost prediction models occurred, and the ensemble CAB-XED framework underwent training and testing utilizing linear regression. Figure 8 illustrates the model's performance in terms of metrics such as MSE, MAE, MAPE, and RMSE during both the training and validation phases, providing a comprehensive overview of the ensemble framework's effectiveness.

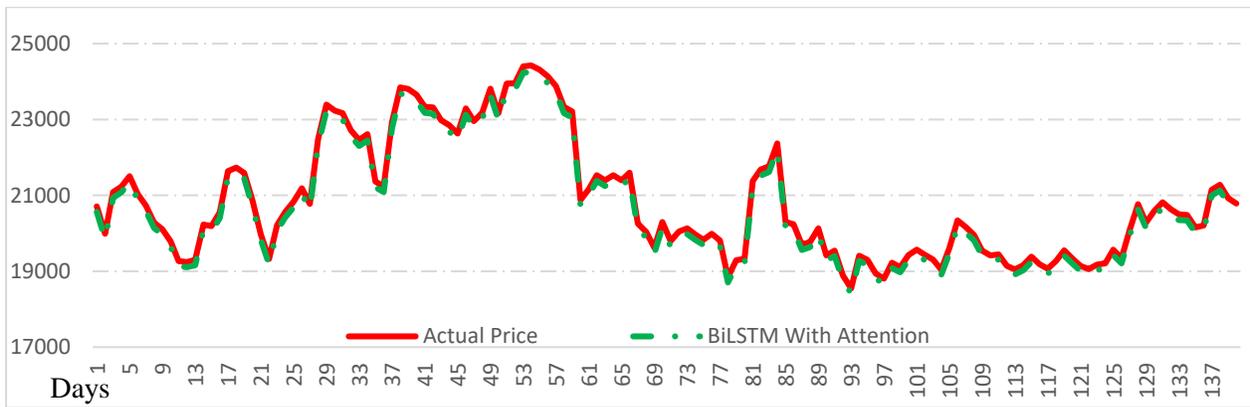

*Figure 9. Bitcoin Price Prediction Using BiLSTM with new attention*

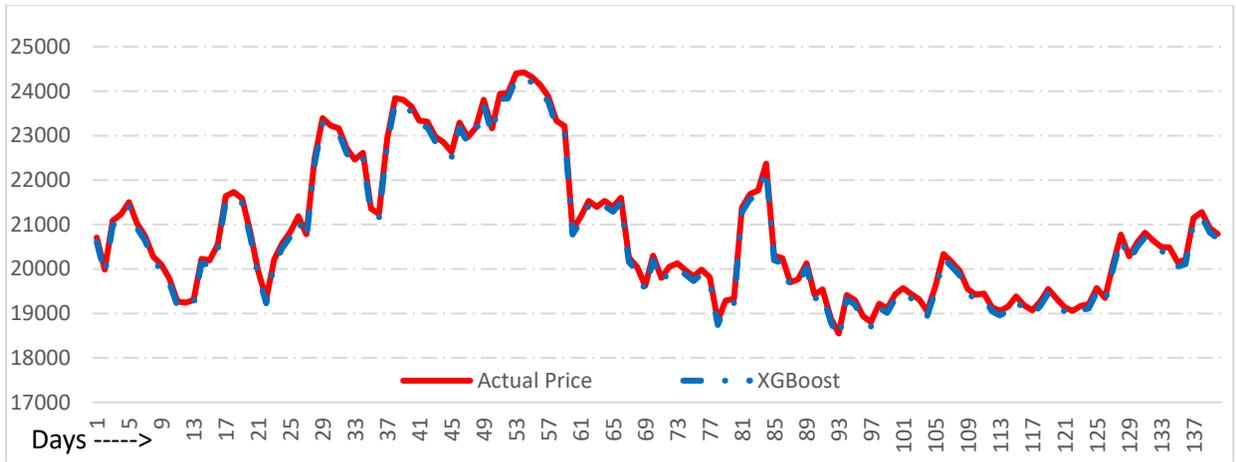

*Figure 10. Bitcoin Price Prediction Using XGBoost*

The test results, outlined in Figure 12, further demonstrate the efficacy of the proposed CAB-XED framework in real-world predictive scenarios. The second dataset (Dataset 2) is used to test the proposed CAB-XED framework. The Dataset 2 was to feed the pre-trained BiLSTM and XGBoost models, and their predictions were subsequently fed into the pre-trained ensemble model of linear regression to obtain the final prediction, as shown in Figure 13.

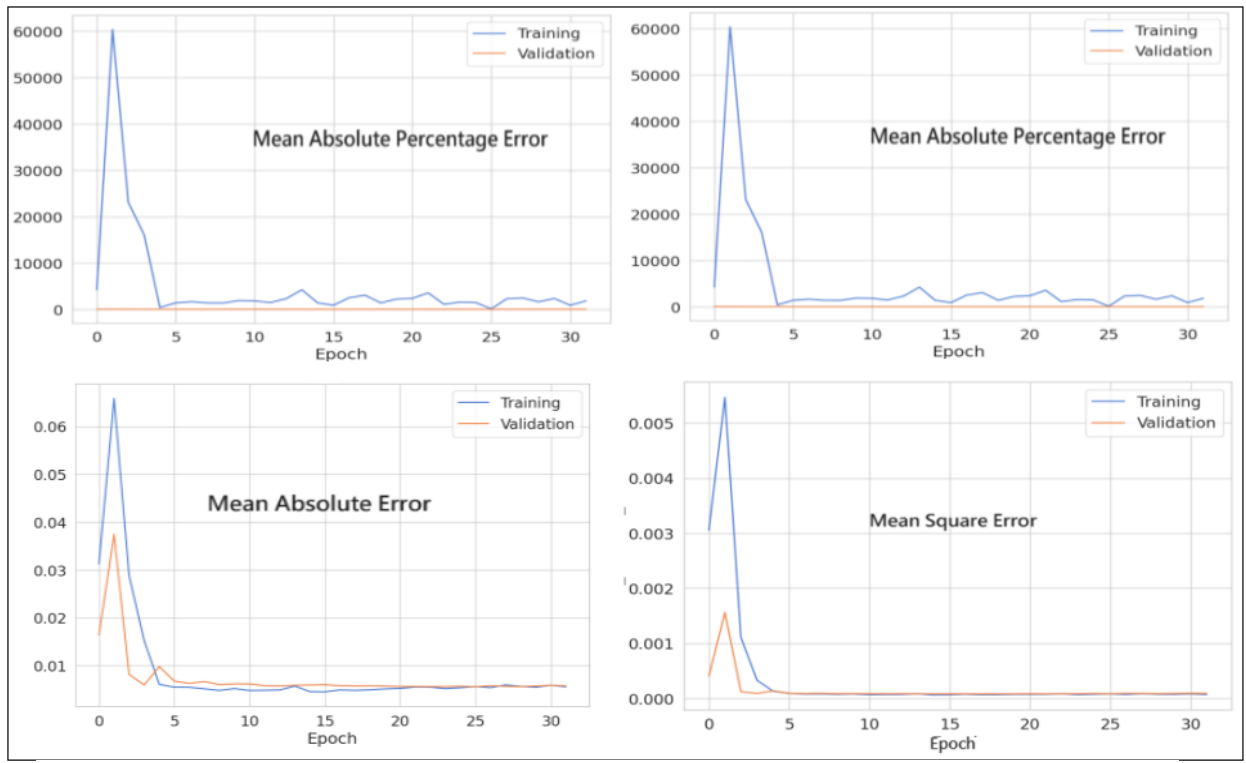

*Figure 11. Proposed CAB-XED framework Training and Performance Evaluation*

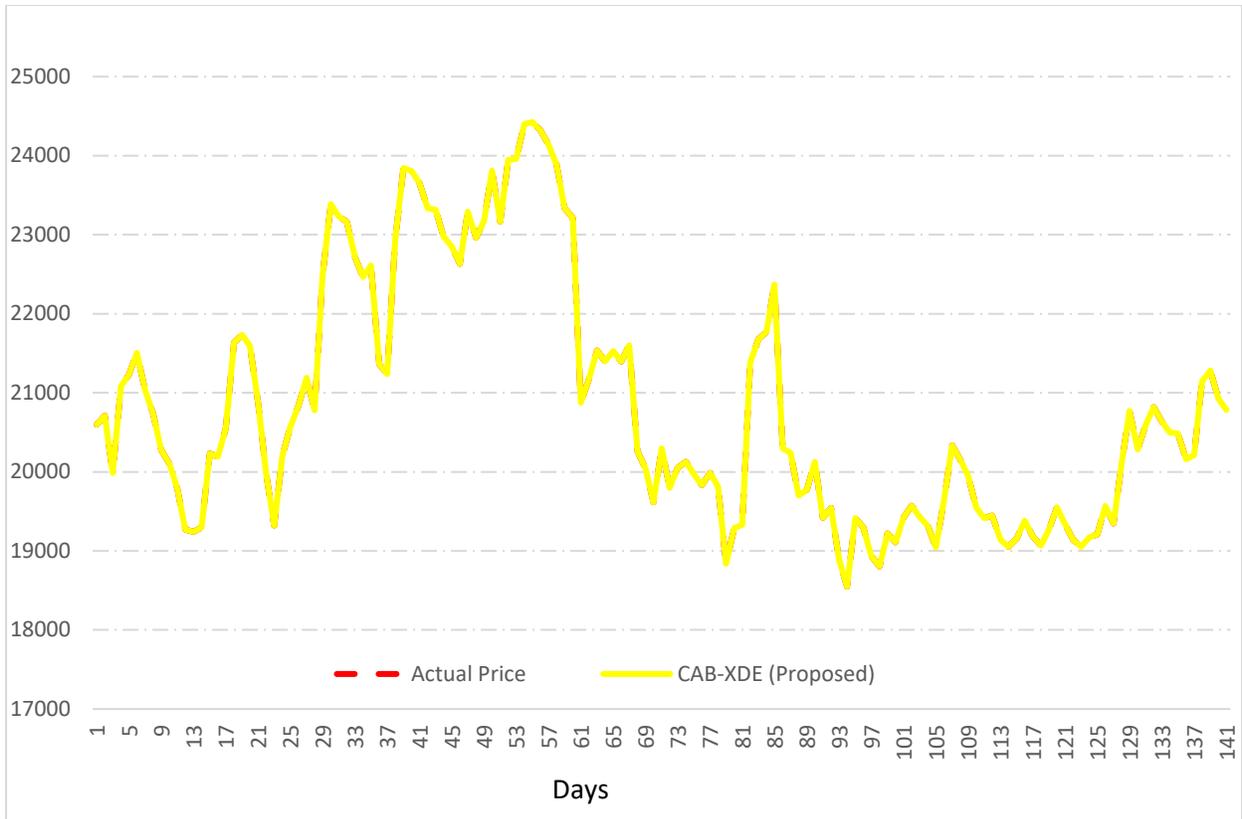

*Figure 12.Training and Validating the proposed CAB-XED framework*

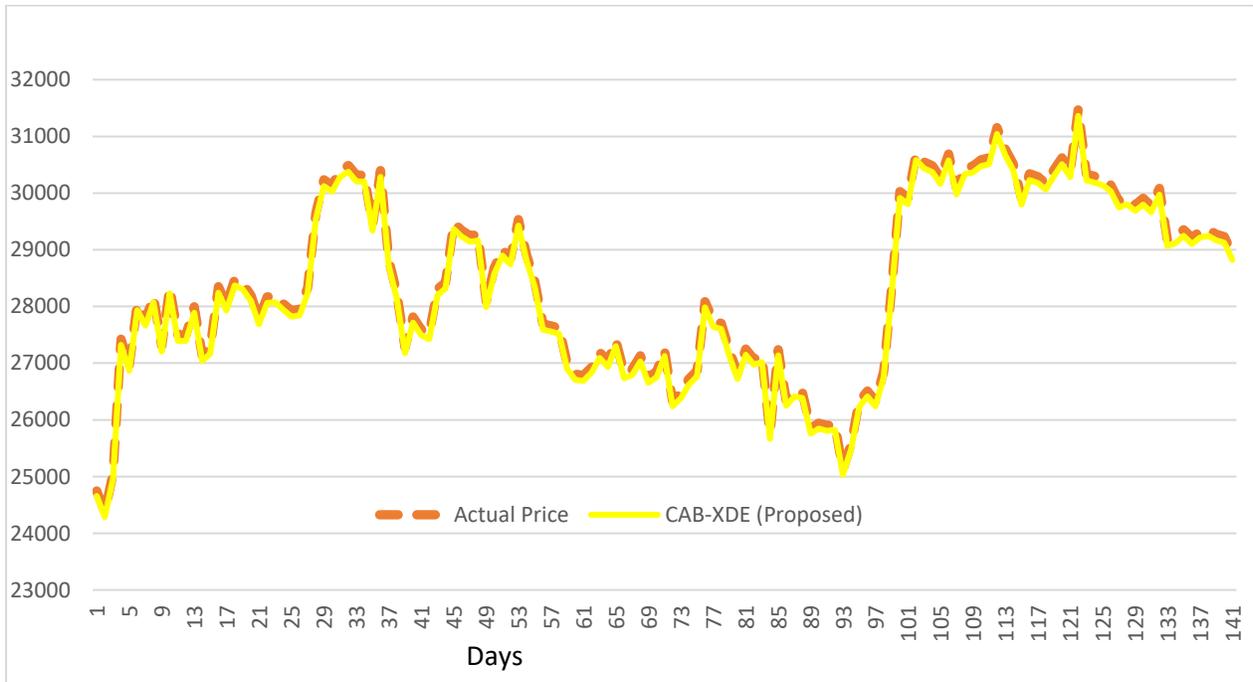

*Figure 13. Testing the proposed CAB-XED framework*

## 5.2 Verification of Combined Model Effectiveness:

The efficacy of the proposed CAB-XED framework is validated by employing LSTM, BiLSTM, attention-LSTM, attention BiLSTM, new attention customized BiLSTM, XGBoost, and the proposed CAB-XDE framework for data prediction. Table 3 offers a comparative analysis of previous studies on Bitcoin across various distributions, while Table 4 presents the Error Analysis of the proposed framework alongside state-of-the-art models. Figure 14 complements these analyses. Additionally, Figure 14 is locally enlarged in Figure 15 to provide a clearer observation of the trend and proximity between each model's forecast results and the actual values. This comparative analysis serves to underscore the performance of the proposed CAB-XED framework in generating predictions. Additionally, we have provided the detailed performance gain of the proposed CAB-XED framework over the existing state of the art models in Figure 16.

*Table 3. Error Analysis of the previous studies employed on Bitcoin at various distributions.*

| Model | MAPE% | MAE | RMSE |
|---|---|---|---|
| LSTM vs. ARIMA[47] | ---- | ----- | 197.515 |
| ARIMA 1 Day [48] | 0.87 | ----- | ----- |
| MICDL [49] | ----- | 170 | 265.05 |
| CNNs + LSTM [50] | 2.35 | 209.89 | 258.31 |
| LSTM 1 Day | | 875.06 | 1203.97 |
| LSTM 3 Day | | 926.96 | 1311.71 |
| LSTM 7 Day | | 1191.05 | 1645.36 |
| CNN 1 Day | | 801.5 | 1107.77 |
| CNN 3 Day | | 1363.85 | 1699.56 |
| CNN 7 Day [51] | | 1197.38 | 1670.93 |
| | | | |

*Table 4. Error Analysis of the proposed framework and evaluated with the state of the art Models.*

| Model | MAPE% | MAE | RMSE |
|---|---|---|---|
| LSTM | 1.122 | 251.52 | 326.37 |
| Attention-LSTM | 1.0 | 217.25 | 287.95 |
| XGBoost | 0.87 | 200.09 | 265.05 |
| BiLSTM | 0.86 | 192.48 | 248.36 |
| **Ablation Study** | | | |
| Attention BiLSTM | 0.70 | 154.23 | 205.47 |
| New attention BiLSTM | 0.51 | 137.72 | 144.73 |
| **Proposed CAB-XDE** | **0.37** | **84.40** | **106.14** |

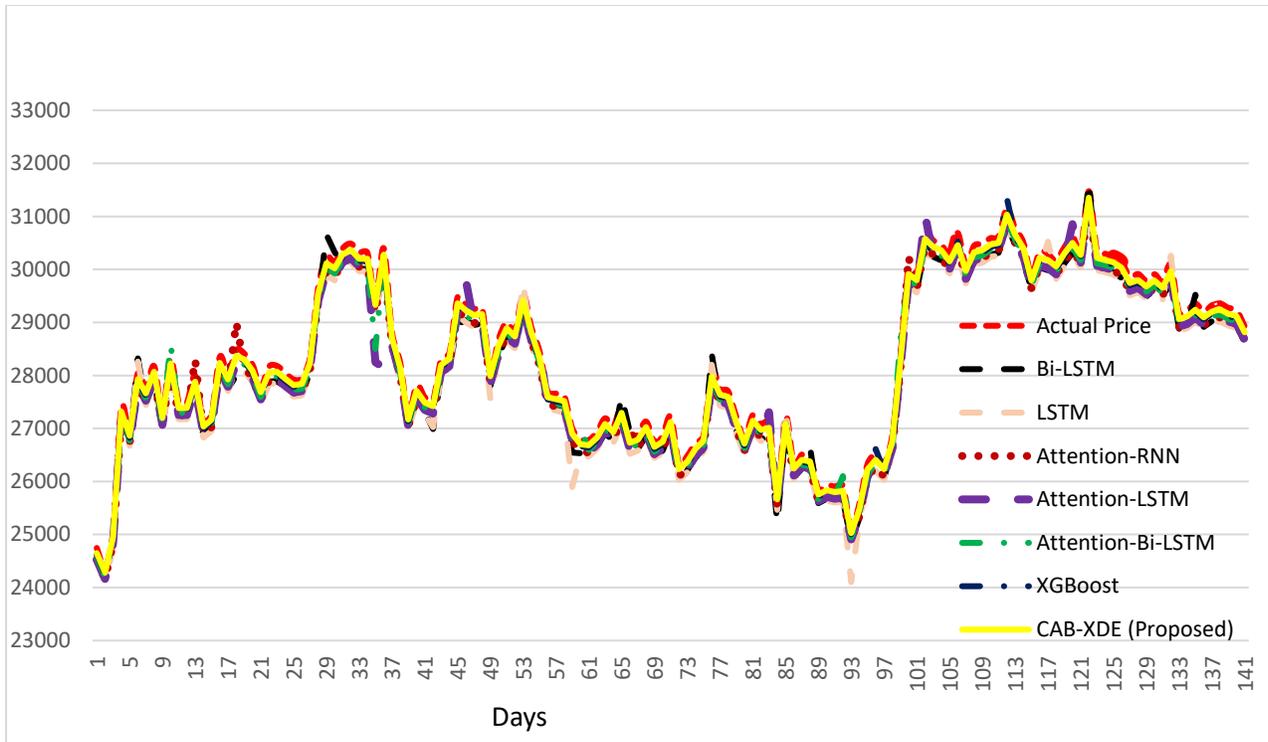

*Figure 14.Comparision of the proposed CAB-XED framework with state of the art models*

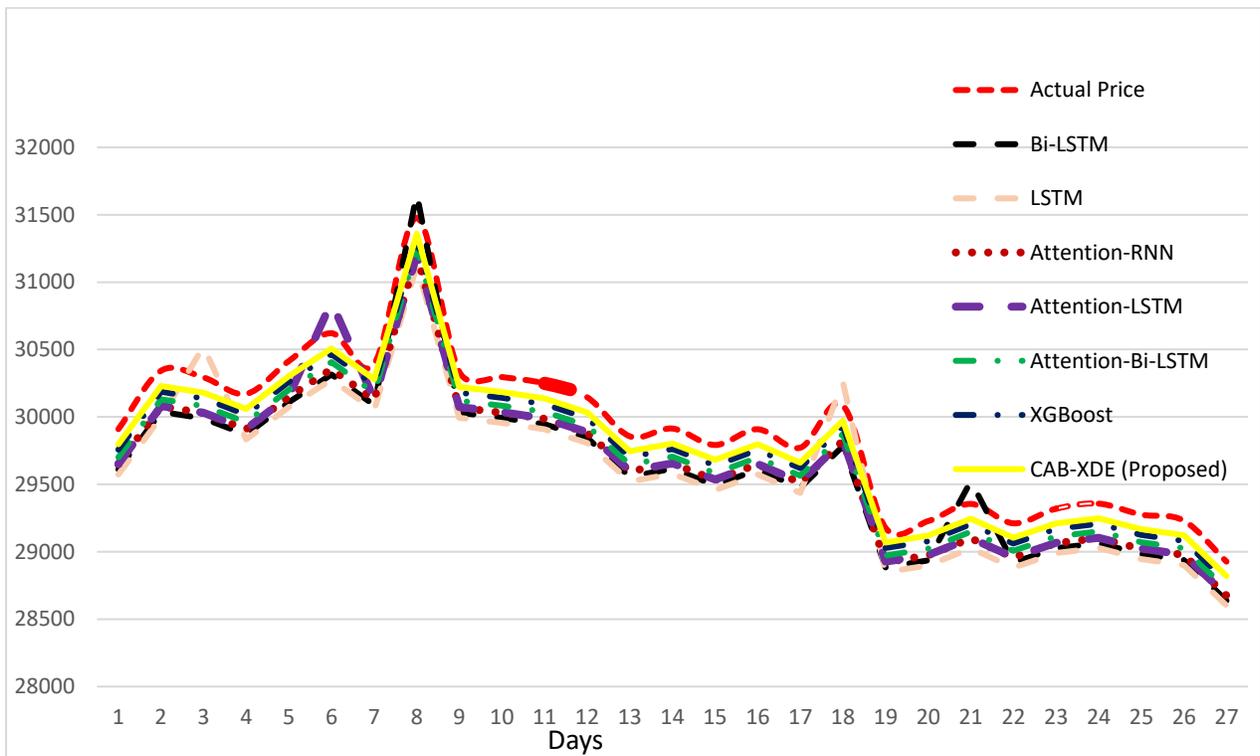

*Figure 15. The subportion of the above Figure 14.*

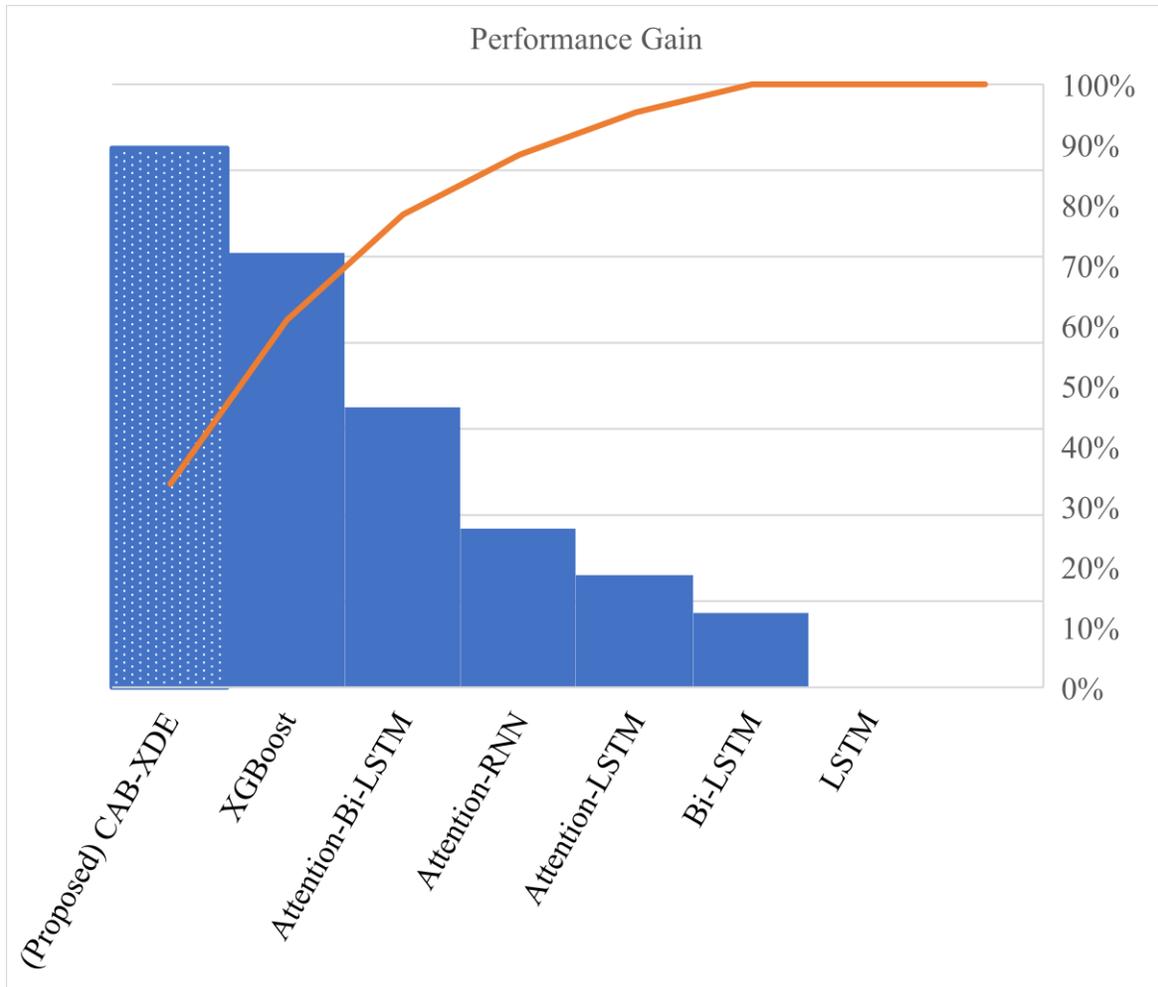

*Figure 16. Proposed CAB-XED framework and state of the art models gain*

### 5.3 Observations and Trends from Results:

1. **Trend Comparison: Predicted Values Compared to Actual Values**

    a) Figure 11 presents a visual comparison between real values and predictions derived from various models, including LSTM, attention-LSTM, BiLSTM, attention BiLSTM, New attention-BiLSTM, XGBoost models, and CAB-XDE. In contrast to the BiLSTM model, which achieves lower accuracy, the proposed CAB-XED framework achieves the best accuracy.

    b) The CAB-XDE's curve closely aligns with the real value curve, showcasing superior fitting effects and a consistent trend.

    c) The CAB-XDE performs exceptionally well in terms of accuracy and sensitivity to changes in proportionality. Figure 11 is locally enlarged in Figure 12 to provide a better understanding of patterns and the proximity between predicted and actual values.

2. **Localized Enlargement (6 July - 01 Aug. 2023):**

Figure 12 highlights the enhanced consistency of the proposed CAB-XED framework's curve with the real value curve, surpassing benchmarks and other models.

The comprehensive and localized figures conclusively demonstrate the better prediction accuracy of proposed CAB-XDE framework compared to other models under consideration.

## 5.4 Comparison and Analysis of Errors

Table 4 outlines MAPE, MAE, and RMSE values for the six models. This analysis unveils valuable insights into their comparative performance.

a) **Model Selection Impact:**The BiLSTM model surpasses LSTM, showcasing lower errors, emphasizing its pivotal role in this study.

b) **attention mechanism Influence:**A thorough model comparison highlights the profound influence of the attention mechanism on prediction accuracy across LSTM, attention-LSTM, BiLSTM, attention-BiLSTM, and new attention customized BiLSTM models.

c) **Benchmark Model Performance:**XGBoost surpasses LSTM and BiLSTM, demonstrating superior prediction with the smallest error. Integrating XGBoost with attention-BiLSTM significantly enhances accuracy..

d) **Combined Model Superiority:**Table 4 data highlights the CAB-XDE's supremacy, presenting minimal MAPE, MAE, and RMSE values at 0.37, 84.40, and 106.14, respectively. This highlights the ensemble approach's effectiveness in minimizing overall prediction errors, surpassing individual models for better accuracy.

## 6. Conclusion

This paper highlights the predictive capabilities of the proposed CAB-XDE framework in refining stock price predictions, particularly within the domain of Bitcoin cryptocurrency. Through meticulous comparative analyses with prominent models, including new attention-BiLSTM, XGBoost, LSTM, BiLSTM, attention LSTM, and attention BiLSTM, key insights are revealed. BiLSTM emerges as a robust solution, adeptly handling complex sequential dependencies, high volatility, and dynamic patterns, while the integration of the attention mechanism significantly enhances its performance, contributing to an overall improvement in prediction accuracy. The incorporation XGBoost proves instrumental, preventing overfitting and enhancing algorithm efficiency. The proposed CAB-XDE framework demonstrates better performance, evidenced by approximately 27.45% lower MAPE, 38.75% lower MAE, and 26.65% lower RMSE compared to state-of-the-art models. Overall, the CAB-XDE framework addresses the limitations of both classical and contemporary forecasting methods, showcasing better performance against state-of-the-art models. Future directions encompass evaluating prediction methods across diverse datasets, expanding the array of evaluation indicators, and optimizing parameters and hyperparameters of models utilizing methodologies like Bayesian optimization. Moreover, incorporating factors such as external influences, legal and regulatory aspects, seasonality trends, as additional input features is crucial. These strategic considerations aim to refine the model, extending its application to diverse fields, thereby broadening its impact and contributing to advancements in predictive modeling across various domains.

**Acknowledgement**

The authors gratefully acknowledge the Artificial Intelligence Lab, Department of Computer Systems Engineering, University of Engineering and Applied Sciences (UEAS), Swat, for providing the necessary resources.

# References

[1] A. Q. Md *et al.*, "Novel optimization approach for stock price forecasting using multi-layered sequential LSTM," *Appl. Soft Comput.*, vol. 134, p. 109830, Feb. 2023, doi: 10.1016/J.ASOC.2022.109830.

[2] M. Khosravi and M. M. Ghazani, "Novel insights into the modeling financial time-series through machine learning methods: Evidence from the cryptocurrency market," *Expert Syst. Appl.*, vol. 234, p. 121012, Dec. 2023, doi: 10.1016/J.ESWA.2023.121012.

[3] "Stock Market Speculation," *Online.* https://study.com/academy/lesson/what-is-speculation-in-the-stock-market.html

[4] L. Almeida and E. Vieira, "Technical Analysis, Fundamental Analysis, and Ichimoku Dynamics: A Bibliometric Analysis," *Risks 2023, Vol. 11, Page 142*, vol. 11, no. 8, p. 142, Aug. 2023, doi: 10.3390/RISKS11080142.

[5] S. B L and S. B R, "Combined deep learning classifiers for stock market prediction: integrating stock price and news sentiments," *Kybernetes*, vol. 52, no. 3, pp. 748–773, Mar. 2023, doi: 10.1108/K-06-2021-0457/FULL/XML.

[6] D. Sheth and M. Shah, "Predicting stock market using machine learning: best and accurate way to know future stock prices," *Int. J. Syst. Assur. Eng. Manag.*, vol. 14, no. 1, pp. 1–18, Feb. 2023, doi: 10.1007/S13198-022-01811-1/METRICS.

[7] B. Gülmez, "Stock price prediction with optimized deep LSTM network with artificial rabbits optimization algorithm," *Expert Syst. Appl.*, vol. 227, p. 120346, Oct. 2023, doi: 10.1016/J.ESWA.2023.120346.

[8] Z. Rauf, A. Sohail, S. H. Khan, A. Khan, J. Gwak, and M. Maqbool, "attention-guided multi-scale deep object detection framework for lymphocyte analysis in IHC histological images," *Microscopy*, vol. 72, no. 1, pp. 27–42, Feb. 2023, doi: 10.1093/JMICRO/DFAC051.

[9] M. M. Zahoor and S. H. Khan, "Brain Tumor MRI Classification using a Novel Deep Residual and Regional CNN," Nov. 2022, Accessed: Oct. 06, 2023. [Online]. Available: https://arxiv.org/abs/2211.16571v2

[10] S. H. Khan, R. Iqbal, and S. Naz, "A Recent Survey of the Advancements in Deep Learning Techniques for Monkeypox Disease Detection," *Inst. Univ. Educ. Física y Deport.*, vol. 9, no. 2, pp. 43–56, Nov. 2, [Online]. Available: https://revistas.udea.edu.co/index.php/viref/article/view/342196/20806106

[11] S. H. Khan *et al.*, "COVID-19 infection analysis framework using novel boosted CNNs and radiological images," *Sci. Rep.*, vol. 13, no. 1, p. 21837, Dec. 2023, doi: 10.1038/s41598-023-49218-7.

[12] S. H. Khan, N. S. Shah, R. Nuzhat, A. Majid, H. Alquhayz, and A. Khan, "Malaria parasite classification framework using a novel channel squeezed and boosted CNN," *Microscopy*, vol. 71, no. 5, pp. 271–282, Oct. 2022, doi: 10.1093/JMICRO/DFAC027.

[13] S. H. Khan *et al.*, "A new deep boosted CNN and ensemble learning based IoT malware detection," *Comput. Secur.*, vol. 133, p. 103385, Oct. 2023, doi: 10.1016/j.cose.2023.103385.

[14] M. Asam, S. H. Khan, T. Jamal, U. Zahoora, and A. Khan, "Malware Classification Using Deep

Boosted Learning," Jul. 2021, Accessed: Oct. 06, 2023. [Online]. Available: http://arxiv.org/abs/2107.04008

[15] S. Qamar, S. H. Khan, M. A. Arshad, M. Qamar, J. Gwak, and A. Khan, "Autonomous Drone Swarm Navigation and Multitarget Tracking With Island Policy-Based Optimization Framework," *IEEE Access*, vol. 10, no. 1, pp. 91073–91091, 2022, doi: 10.1109/ACCESS.2022.3202208.

[16] M. A. Arshad *et al.*, "Drone Navigation Using Region and Edge Exploitation-Based Deep CNN," *IEEE Access*, vol. 10, no. 1, pp. 95441–95450, 2022, doi: 10.1109/ACCESS.2022.3204876.

[17] A. H. Khan *et al.*, "A performance comparison of machine learning models for stock market prediction with novel investment strategy," *PLoS One*, vol. 18, no. 9, p. e0286362, Sep. 2023, doi: 10.1371/journal.pone.0286362.

[18] Y. Zhao and G. Yang, "Deep Learning-based Integrated Framework for stock price movement prediction," *Appl. Soft Comput.*, vol. 133, p. 109921, Jan. 2023, doi: 10.1016/J.ASOC.2022.109921.

[19] G. Sonkavde, D. S. Dharrao, A. M. Bongale, S. T. Deokate, D. Doreswamy, and S. K. Bhat, "Forecasting Stock Market Prices Using Machine Learning and Deep Learning Models: A Systematic Review, Performance Analysis and Discussion of Implications," *Int. J. Financ. Stud.*, vol. 11, no. 3, p. 94, Jul. 2023, doi: 10.3390/ijfs11030094.

[20] A. Khan *et al.*, "A Recent Survey of Vision Transformers for Medical Image Segmentation," *Mass Commun. Soc.*, vol. 3, no. 10, pp. 349–83, Dec. 2023, doi: 10.1163/_q3_SIM_00374.

[21] S. H. Khan *et al.*, "COVID-19 detection and analysis from lung CT images using novel channel boosted CNNs," *Expert Syst. Appl.*, vol. 229, p. 120477, Nov. 2023, doi: 10.1016/J.ESWA.2023.120477.

[22] S. H. Khan, "Malaria Parasitic Detection using a New Deep Boosted and Ensemble Learning Framework," Dec. 2022, doi: 10.1088/1674-1137/acb7ce.

[23] S. H. Khan, A. Khan, Y. S. Lee, M. Hassan, and W. K. Jeong, "Segmentation of shoulder muscle MRI using a new Region and Edge based Deep Auto-Encoder," *Multimed. Tools Appl.*, vol. 82, no. 10, pp. 14963–14984, Apr. 2023, doi: 10.1007/S11042-022-14061-X/METRICS.

[24] U. Zahoora, A. Khan, M. Rajarajan, S. H. Khan, M. Asam, and T. Jamal, "Ransomware detection using deep learning based unsupervised feature extraction and a cost sensitive Pareto Ensemble classifier," *Sci. Rep.*, vol. 12, no. 1, p. 15647, Sep. 2022, doi: 10.1038/s41598-022-19443-7.

[25] S. Qamar, S. H. Khan, M. A. Arshad, M. Qamar, and A. Khan, "Autonomous Drone Swarm Navigation and Multi-target Tracking in 3D Environments with Dynamic Obstacles," Feb. 2022, Accessed: Oct. 06, 2023. [Online]. Available: http://arxiv.org/abs/2202.06253

[26] A. EL ZAAR, N. BENAYA, T. BAKIR, A. MANSOURI, and A. EL ALLATI, "Prediction of US 30-years-treasury-bonds mouvement and trading entry point using Robust 1DCNN-BiLSTM-XGBoost algorithm," *Authorea Prepr.*, Apr. 2023, doi: 10.22541/AU.168079685.52841217/V1.

[27] P. Chhajer, M. Shah, and A. Kshirsagar, "The applications of artificial neural networks, support

vector machines, and long–short term memory for stock market prediction," *Decis. Anal. J.*, vol. 2, p. 100015, Mar. 2022, doi: 10.1016/J.DAJOUR.2021.100015.

[28] A. A, R. R, V. R. S, and A. M. Bagde, "Predicting Stock Market Time-Series Data using CNN-LSTM Neural Network Model," May 2023, [Online]. Available: https://arxiv.org/pdf/2305.14378

[29] A. John and T. Latha, "Stock market prediction based on deep hybrid RNN model and sentiment analysis," *Automatika*, vol. 64, no. 4, pp. 981–995, Oct. 2023, doi: 10.1080/00051144.2023.2217602.

[30] J. Zhang, L. Ye, and Y. Lai, "Stock Price Prediction Using CNN-BiLSTM-attention Model," *Math. 2023, Vol. 11, Page 1985*, vol. 11, no. 9, p. 1985, Apr. 2023, doi: 10.3390/MATH11091985.

[31] J. Shah, D. Vaidya, and M. Shah, "A comprehensive review on multiple hybrid deep learning approaches for stock prediction," *Intell. Syst. with Appl.*, vol. 16, p. 200111, Nov. 2022, doi: 10.1016/J.ISWA.2022.200111.

[32] Y. Guo, J. Guo, B. Sun, J. Bai, and Y. Chen, "A new decomposition ensemble model for stock price forecasting based on system clustering and particle swarm optimization," *Appl. Soft Comput.*, vol. 130, p. 109726, Nov. 2022, doi: 10.1016/J.ASOC.2022.109726.

[33] C. Cui, P. Wang, Y. Li, and Y. Zhang, "McVCsB: A new hybrid deep learning network for stock index prediction," *Expert Syst. Appl.*, vol. 232, p. 120902, Dec. 2023, doi: 10.1016/J.ESWA.2023.120902.

[34] S. Zhang, J. Luo, S. Wang, and F. Liu, "Oil price forecasting: A hybrid GRU neural network based on decomposition–reconstruction methods," *Expert Syst. Appl.*, vol. 218, p. 119617, May 2023, doi: 10.1016/J.ESWA.2023.119617.

[35] T. Xu and X. He, "EMD-BiLSTM Stock Price Trend Forecasting Model based on Investor Sentiment," *Front. Comput. Intell. Syst.*, vol. 4, no. 3, pp. 139–143, Jul. 2023, doi: 10.54097/fcis.v4i3.11268.

[36] S. ZHAO, X. LIN, and X. WENG, "attention-BiLSTM stock price trend prediction model based on empirical mode decomposition and investor sentiment," *J. Comput. Appl.*, vol. 43, no. S1, p. 112, Jun. 2023, doi: 10.11772/J.ISSN.1001-9081.2022060863.

[37] K. Ateeq, A. A. Al Zarooni, A. Rehman, and M. A. Khan, "A Mechanism for Bitcoin Price Forecasting using Deep Learning," *Int. J. Adv. Comput. Sci. Appl.*, vol. 14, no. 8, pp. 441–448, 2023, doi: 10.14569/IJACSA.2023.0140849.

[38] J. Chen, "Analysis of Bitcoin Price Prediction Using Machine Learning," *J. Risk Financ. Manag. 2023, Vol. 16, Page 51*, vol. 16, no. 1, p. 51, Jan. 2023, doi: 10.3390/JRFM16010051.

[39] S. Ranjan, P. Kayal, and M. Saraf, "Bitcoin Price Prediction: A Machine Learning Sample Dimension Approach," *Comput. Econ.*, vol. 61, no. 4, pp. 1617–1636, Apr. 2023, doi: 10.1007/S10614-022-10262-6/TABLES/37.

[40] P. L. Seabe, C. R. B. Moutsinga, and E. Pindza, "Forecasting Cryptocurrency Prices Using LSTM, GRU, and Bi-Directional LSTM: A Deep Learning Approach," *Fractal Fract. 2023, Vol. 7, Page 203*, vol. 7, no. 2, p. 203, Feb. 2023, doi: 10.3390/FRACTALFRACT7020203.

[41]    Z. Jiang, "An attention GRU-XGBoost Model for Stock Market Prediction Strategies," *ACM Int. Conf. Proceeding Ser.*, Nov. 2022, doi: 10.1145/3573834.3573837/ASSETS/HTML/IMAGES/IMAGE14.PNG.

[42]    Y. Du and J. Wu, "No One Left Behind: Improving the Worst Categories in Long-Tailed Learning," Mar. 2023, [Online]. Available: http://arxiv.org/abs/2303.03630

[43]    M. Maiti, "Dynamics of bitcoin prices and energy consumption," *Chaos, Solitons Fractals X*, vol. 9, p. 100086, Dec. 2022, doi: 10.1016/J.CSFX.2022.100086.

[44]    "Power of Recurrent Neural Networks (RNN): Revolutionizing AI." https://www.simplilearn.com/tutorials/deep-learning-tutorial/rnn (accessed Dec. 02, 2023).

[45]    J. Lou, L. Cui, and Y. Li, "Bi-LSTM Price Prediction based on attention mechanism," *arXiv.org*, 2022, doi: 10.48550/ARXIV.2212.03443.

[46]    S. Ranjan, P. Kayal, and M. Saraf, "Bitcoin Price Prediction: A Machine Learning Sample Dimension Approach," *Comput. Econ.*, vol. 61, no. 4, pp. 1617–1636, Apr. 2023, doi: 10.1007/s10614-022-10262-6.

[47]    C. M. Liapis, A. Karanikola, and S. Kotsiantis, "Investigating Deep Stock Market Forecasting with Sentiment Analysis," *Entropy 2023, Vol. 25, Page 219*, vol. 25, no. 2, p. 219, Jan. 2023, doi: 10.3390/E25020219.

[48]    I. M. Wirawan, T. Widiyaningtyas, and M. M. Hasan, "Short Term Prediction on Bitcoin Price Using ARIMA Method," in *2019 International Seminar on Application for Technology of Information and Communication (iSemantic)*, IEEE, Sep. 2019, pp. 260–265. doi: 10.1109/ISEMANTIC.2019.8884257.

[49]    I. E. Livieris, N. Kiriakidou, S. Stavroyiannis, and P. Pintelas, "An Advanced CNN-LSTM Model for Cryptocurrency Forecasting," *Electronics*, vol. 10, no. 3, p. 287, Jan. 2021, doi: 10.3390/electronics10030287.

[50]    Y. Li and W. Dai, "Bitcoin price forecasting method based on CNN-LSTM hybrid neural network model," *J. Eng.*, vol. 2020, no. 13, pp. 344–347, Jul. 2020, doi: 10.1049/joe.2019.1203.

[51]    N. S. Wen and L. S. Ling, "Evaluation of Cryptocurrency Price Prediction Using LSTM and CNNs Models," *JOIV Int. J. Informatics Vis.*, vol. 7, no. 3–2, pp. 2016–2024, Nov. 2023, doi: 10.30630/joiv.7.3-2.2344.